\documentclass[apj,numberedappendix]{emulateapj}

\slugcomment{Accepted to ApJ}

\shorttitle{On the Evolution of the Cores and Lobes}
\shortauthors{Yuan \& Wang}

\begin{document}

\title{On the Evolution of the Cores of Radio Sources and Their Extended Radio Emission}

%
\author{
Zunli Yuan\altaffilmark{1,2,3}, Jiancheng Wang\altaffilmark{1,2}
}

\altaffiltext{1} {National Astronomical Observatories, Yunnan Observatory, Chinese Academy
of Sciences,  Kunming 650011, China}
\altaffiltext{2} {Key Laboratory for the Structure and Evolution of Celestial Objects,
Chinese Academy of Sciences,  Kunming 650011, China}
\altaffiltext{3} {Graduate School, Chinese Academy of Sciences, Beijing, P.R. China}

\email{yuanzunli@ynao.ac.cn}

\begin{abstract}
The work in this paper aims at determining the evolution and possible co-evolution of radio-loud active galactic nuclei (AGNs) and their cores via their radio luminosity functions (i.e., total and core RLF respectively). Using a large combined sample of 1063 radio-loud AGNs selected at low radio frequency, we investigate the radio luminosity function (RLF) at 408 MHz of steep-spectrum radio sources. Our results support a luminosity-dependent evolution. Using core flux density data of the complete sample 3CRR, we investigate the core RLF at 5.0 GHz. Based on the combined sample with incomplete core flux data, we also estimate the core RLF using a modified factor of completeness. Both results are consistent and show that the comoving number density of radio cores displays a persistent decline with redshift, implying a negative density evolution. We find that the core RLF is obviously different from the total RLF at 408 MHz band which is mainly contributed by extended lobes, implying that the cores and extended lobes could not be co-evolving at radio emission.
\end{abstract}


\section{Introduction}

The radio-loud AGNs are powerful objects in which strong radio emission originate from non-thermal processes associated with the bipolar outflow from central active nuclei. In the observation of low radio frequency, they mainly consist of the radio galaxies (RGs) and steep-spectrum radio quasars (SSRQs), such as GHz Peaked Spectrum sources (GPSs), Compact Steep Spectrum sources (CSSs), Giant Radio Galaxies (GRGs) etc. Some radio galaxies have linear sizes as large as several megaparsecs, and are possibly the largest individual objects in the Universe \citep{b27}. The radio structures of radio-loud AGNs often show various morphologies, but a two-component structure with core and lobes (generally showing a steep-spectrum) is usually in common. Traditionally, the `core' is defined as a component which is normally unresolved on arcsecond scales and has a flat spectrum. It is usually believed that the core is coincident with the active nucleus that transports energy and matter to the extended lobes by the outflow. \citet{b15} found a statistically significant correlation between the core radio power at 5.0 GHz and the total radio power at 408 MHz which is mainly dominated by extended lobes. Similar relations were also found by other authors \citep[e.g.][]{b17,b44}. In recent years, Hardcastle and his collaborators have found more relations between the properties of cores and extended lobes based on 3CRR radio galaxies/QSOs \citep[see][]{b18,b35,b19}. It is known that the luminosity of radio galaxies/QSOs evolve with time \citep{b38,b3}. As the most active parts in radio AGNs, it is not known whether the cores co-evolve with their parent radio sources. To investigate this problem, we study their radio luminosity functions (i.e., core and total RLF respectively).

The RLF, presenting the space density of radio sources along with redshift and luminosity, is very important because its shape and evolution provide constraints on the nature of radio activity and the cosmic evolution of radio galaxies/QSOs \citep{b43}. Many researches on the total RLF have been done \citep[e.g.][]{b8,b11,b42,b23,b16} since \citet{L1966} performed some precursory work. \citet{b11} found a decline of the RLF beyond $z\thicksim 2.5$, e.g. so-called redshift cut-off. \citet{b8} calculated the RLF of cores based on a sample of 54 bright elliptical galaxies identified with radio sources of the B2 catalogue. However, they did not present any evolution of radio cores with redshift. After then, the research on the RLF of the cores is rare. Because the cores are often unresolved and variable, it is not easy to establish a desirable sample of cores with reliable radio flux. In recent years, the fluxes of cores in many sources were obtained by Very Large Array (VLA), Australia Telescope Compact Array (ATCA), Parkes-Tidbinbilla real-time interferometer (PTI), and Very Long Baseline Interferometry (VLBI). In this paper, we establish a large combined sample of radio galaxies/QSOs by collecting data from the literature. Many sources have 5 GHz core flux (57.1\%), total flux at 408 MHz (100\%) and redshifts (97.0\%). We then derive the core RLF at 5 GHz and total RLF at 408 MHz respectively, and compare their properties to study the co-evolution in detail.

The sources in our sample are all steep-spectrum ones which mainly consist of radio galaxies and steep-spectrum QSOs. The flat-spectrum counterparts of these sources are flat-spectrum QSOs and BL Lacs. In the light of Unified Scheme of AGNs \citep[e.g.][]{A93,U95}, the steep-spectrum sources are inclined at larger angles to the observer. This `edge-on' character make it possible to observe the extended lobes and radio cores at the same time. In addition, the radio cores in steep-spectrum sources are less affected by Doppler-boosting compared with that in flat-spectrum sources. Therefore we use steep-spectrum sources to study the evolution relationship between the cores and their extended radio emission.

Throughout the paper, we adopt a Lambda Cold Dark Matter cosmology with the parameters $\Omega_{m}$ = 0.27,  $\Omega_{\Lambda}$ = 0.73, and $H_{0}$ = 71 km s$^{-1}$ Mpc$^{-1}$.

\section[]{The combined sample}

In general, to construct the RLF of radio sources, a complete sample consisting of every radio sources in a certain area on the sky brighter than a specified flux limit at the selection frequency is required. A set of independent complete samples in different surveys with different flux limit can also be combined together into one `coherent' sample \citep{b1}. In the paper we construct such a coherent sample combined with four independent complete samples (described below) totalling 1063 sources. These samples are selected from low-frequency survey, e.g. one sample at 178 MHz and three at 408 MHz. They have smaller orientation bias \citep{b18} than high-frequency selected samples. The following sections present detailed information about the sub-samples. The final list of radio galaxies/QSOs can be found in the Appendix A.

\subsection{The 3CRR bright sample}

The 3CRR bright sample, also known as LRL sample \citep{b26}, is a sample of 173 sources at 178 MHz with flux-density above 10.9 Jy, over a 4.233 sr region of the sky. Assuming a typical index of 0.8, we obtain the flux limit to be 5.6 Jy at 408 MHz. Following \citet{b43}, we exclude three sources, 3C231, 3C345, 3C454.3, in which 3C231 (M82) is a starburst galaxy, while 3C345 and 3C454.3 actually have fluxes lower than the flux limit. This sample thus includes 170 sources, in which 169 sources have core flux densities at 5 GHz and all sources have spectroscopic redshifts.

\subsection{The MRC 1-Jy sample}

\begin{figure}
  \centerline{
    \includegraphics[scale=0.46,angle=0]{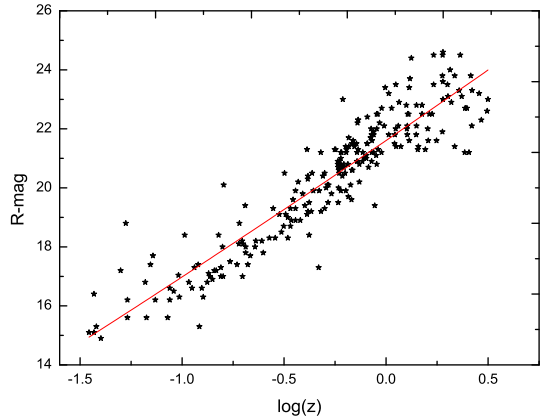}}
  \caption[R-z relation]{\label{Photoz}R-z relation from MRC 1-Jy data (McCarthy et al. 1996) for extended sources. The stars represent objects with optical ID and spectroscopic redshift information. The red solid line shows the quadratic fit ($R = 21.60 + 4.72 \log z + 0.10 {(\log z)}^2$) from which photometric redshifts are estimated.}
\end{figure}

\begin{figure}
  \centerline{
    \includegraphics[scale=0.46,angle=0]{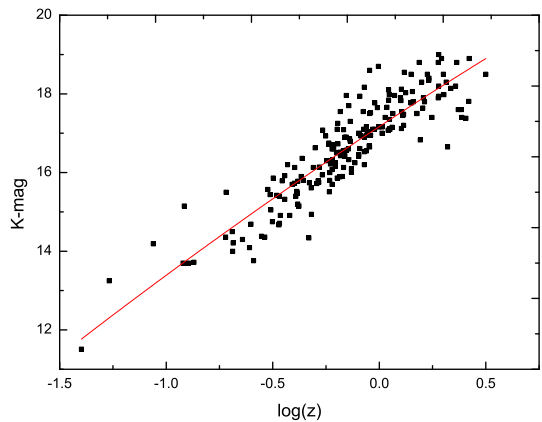}}
  \caption[R-z relation]{\label{Photoz}K-z relation for extended sources of MRC 1-Jy. The K-band magnitudes information are from McCarthy (by private communication). The squares represent objects with optical ID and spectroscopic redshift information. The red solid line shows the quadratic fit ($K = 17.17 + 3.57 \log z - 0.21 {(\log z)}^2$) from which photometric redshifts are estimated.}
\end{figure}

The MRC 1-Jy sample was selected by McCarthy and his collaborators from the Molonglo Reference Catalog \citep[MRC][]{b29,b28} at 408 MHz \citep[see][for details]{b33}. The sample consists of 559 radio sources, in which 111 ones are quasars \citep{b24} and others are radio galaxies \citep{b25}. The flux limit is 0.95 Jy at 408 MHz and the sky area covered is 0.56 sr. 18 sources are excluded because they are flat-spectrum ones with spectral index near 408 MHz $\alpha<0.5$ ($S=\nu^{-\alpha}$). For the left 541 sources, core flux densities of 207 objects were retrieved from the literature. Only 375 (69.3\%) sources have spectroscopic redshift data. Since the main sources without redshifts are radio galaxies, we estimate their photometric redshifts using the empirical relation of $R-z$ or $K-z$ derived from extended sources in the MRC 1-Jy sample with galaxy identifications. The R-band magnitude data are from \citet{b33} and K-band data are from P. J. McCarthy (by private communication). The two relations are shown in Fig. 1 and Fig. 2 respectively. The best fits to these data are
\begin{eqnarray}
R = 21.60 + 4.72 \log z + 0.10 {(\log z)}^2
\end{eqnarray}
\begin{eqnarray}
K = 17.17 + 3.57 \log z - 0.21 {(\log z)}^2
\end{eqnarray}

If the redshift can be estimated using both relations, we take their average value as $z_{est}$. Photometric redshifts of 138 sources are estimated, bringing the total number of sources with redshift information to 513 (94.8\%).

\subsection{The MS4 sample}

The Molonglo Southern 4 Jy sample (hereafter MS4) was selected by \citet{b5,b6} from the MRC at 408 MHz. Located in a different region of the sky with the MRC 1-Jy sample, MS4 has a flux limit of 4.0 Jy at 408 MHz and covers a sky area of 2.43 sr. After excluding 10 flat-spectrum sources, there are 218 ones left, in which 213 have spectroscopic redshifts. The core fluxes of 160 sources at 5 GHz are retrieved from the literature.

\subsection{The BRL sample}

This sample, selected by \citet{b2}, was also from the MRC at 408 MHz. It locates in different sky from the MS4 sample and partly overlaps with the MRC 1-Jy sample. The flux limit of BRL is 5.0 Jy at 408 MHz and the sky area is 3.606 sr. The sample contains 178 sources with spectroscopic redshifts. Removing the sources repeated in MRC 1-Jy sample, as well as 9 flat-spectrum ones, we finally obtain 134 sources in the sample, in which 65 objects have the core fluxes at 5 GHz.

\subsection{The ratio of core to extended radio fluxes}

Although all the 1063 sources in our combined sample are steep-spectrum, this can not guarantee that the Doppler boosting effects are negligible. In order to estimate these effects on RLF calculation, we have evaluated the ratio of core to extended radio fluxes for our sample. In the literature, this ratio is often referred to as the $R_{c}$-parameter \citep{b44}, or as the orientation parameter \citep{1992P}. We calculate the ratio via:

\begin{eqnarray}
R_{c}=\frac{S_{core(5 GHz)}}{S_{total(5 GHz)}-S_{core(5 GHz)}}
\end{eqnarray}

The total fluxes at 5 GHz for most of the sources are retrieved from the literature. For few sources in which no $S_{total(5 GHz)}$ is available, we convert the flux of nearest frequency. In Fig. 3 we show the histograms of the $R_{c}$-parameters for our sample. \citet{1992P} showed that steep-spectrum quasars have their radio axes within $14^{\circ} \lesssim \theta \lesssim 40^{\circ}$, and high-luminosity RGs are in the range $\theta \gtrsim 40^{\circ}$. They associated the $R_{c}$-parameter with $\theta$, i.e., the larger $R_{c}$ corresponds to smaller view angle $\theta$. It is shown in Fig. 3 that 94\% sources have $\log R_{c} <0$, and 70\% sources with $\log R_{c} <-1$. This implies that only few sources ($\thicksim 5\%$) have viewing angles typically smaller $40^{\circ}$. We thus argue that the Doppler boosting effects are not important for RLF calculation.

\begin{figure}
  \centerline{
    \includegraphics[scale=0.46,angle=0]{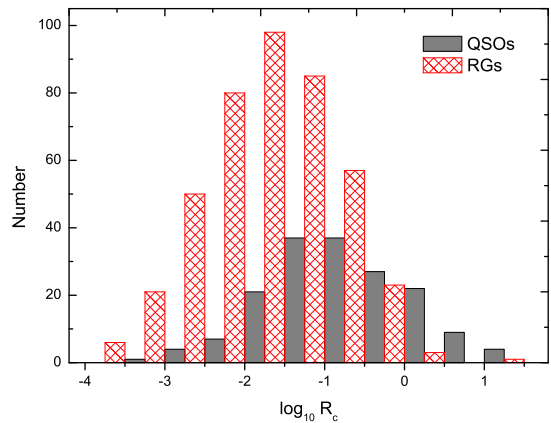}}
  \caption[redshift distribution]{\label{Photoz} The distribution of $R_{c}$-parameter for our sample. The QSO and RG samples are represented by the gray solid and red cross-hatched columns.}
\end{figure}

\section[]{The radio luminosity function}

The RLF $\rho(L,z)$ is defined as the number density of radio sources per unit comoving volume per unit logarithmic luminosity. The luminosity used here is rest-frame monochromatic luminosity (W/Hz). We chose to evaluate the total radio luminosity at 408 MHz to minimize the potential impact of beaming, and also because it is the frequency of the MRC samples. For the 3CRR sources, the fluxes at 408 MHz  are taken from literature or converted from the fluxes at 178 MHz. The core radio luminosity is evaluated at 5 GHz, and is less contaminated by the lobe radiation.

There are many methods to determine the RLF. The maximum likelihood \citep{b32} and binned $1/V_{a}$ methods \citep{b1} are commonly used. In this paper, we use the $1/V_{a}$ method to calculate the total and core RLFs.

\subsection{The binned $1/V_{a}$ methods}

The binned $1/V_{a}$ method, a generalized version of the original $1/V_{max}$ method \citep{b36}, is often used to estimate the luminosity function (LF) of combined complete samples. The method of estimating the LF evolution with the redshift was developed by \citep{b12,b13}.

The RLF in the ranges [$z_{1},z_{2}$] and [$\log_{10}L_{1},\log_{10}L_{2}$] is estimated as
\begin{eqnarray}
\rho(L,z)=(\Delta\log_{10}L)^{-1}\sum_{i=1}^{N}\frac{1}{V_{a}^{i}}
\end{eqnarray}
and the error on each bin \citep{b31} is given by
\begin{eqnarray}
\sigma=(\Delta\log_{10}L)^{-1}[\sum_{i=1}^{N}(V_{a}^{i})^{-2}]^{0.5}.
\end{eqnarray}
$V_{a}^{i}$ is the total accessible volume of radio source $i$ given by
\begin{eqnarray}
V_{a}^{i}=\sum_{j=1}^{M} V_{ij}=C_{j}\Omega_{j}\int_{z_{min}^{ij}}^{z_{max}^{ij}}\frac{dV}{dz}dz,
\end{eqnarray}
where $V_{ij}$ is the accessible volume of the radio source $i$ in sub-sample $j$. $\Omega_{j}$ is the solid angle subtended by sub-sample $j$ and $C_{j}$ is a modified factor of sample completeness. $z_{min}^{ij}$ and $z_{max}^{ij}$ are the minimum and maximum redshifts defined as \citep{b12}
\begin{eqnarray}
z_{min}^{ij}=\max[z_{1},z(L_{i},S_{1j})],
\end{eqnarray}
and
\begin{eqnarray}
z_{max}^{ij}=\min[z_{2},z(L_{i},S_{2j})]
\end{eqnarray}
where $S_{1j}$ and $S_{2j}$ are the minimum and maximum fluxes of sub-sample $j$ respectively.
$z(L,S)$ is the redshift of object with luminosity $L$ and flux $S$. The differential comoving volume is given by \citep{b21}

\begin{eqnarray}
\frac{dV}{dz}=\frac{c}{H_{0}}\frac{d_{L}^{2}[\Omega_{M}(1+z)^{3}+\Omega_{\Lambda}]^{-1/2}}{(1+z)^{2}},
\end{eqnarray}
where $d_{L}$ is the luminosity distance defined as
\begin{eqnarray}
d_{L}=\frac{c}{H_{0}}(1+z)\int_{0}^{z}\frac{dz}{\sqrt{\Omega_{M}(1+z)^{3}+\Omega_{\Lambda}}}.
\end{eqnarray}

\begin{figure*}
  \centerline{
    \includegraphics[scale=0.80,angle=0]{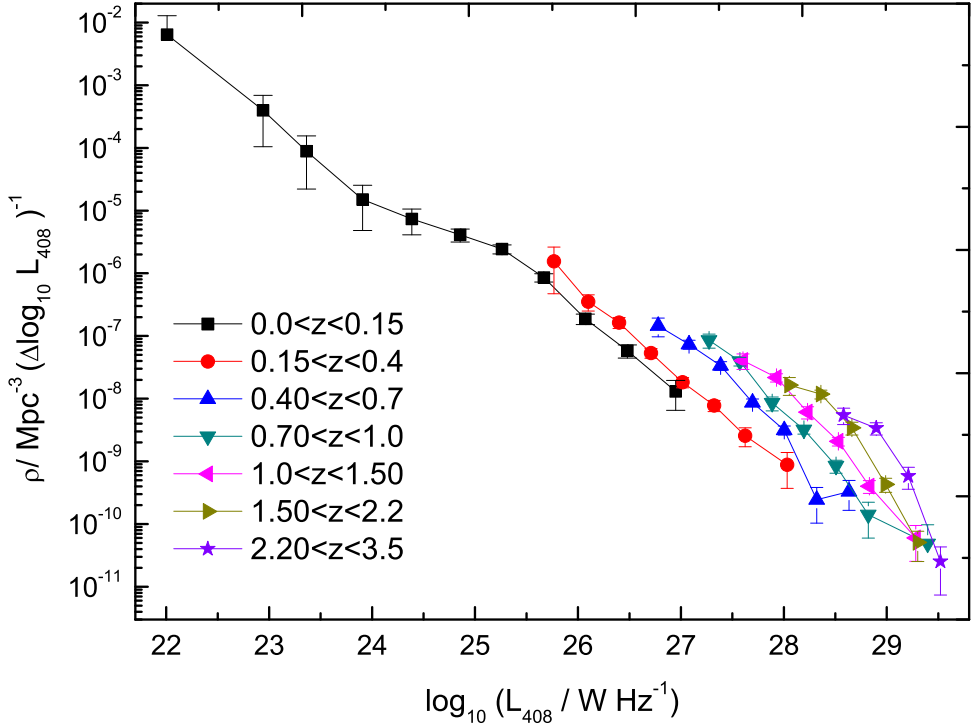}}
  \caption[R-z relation]{\label{Photoz}The binned total RLFs are plotted with Poisson errors from Eq.(5). Different symbols correspond to different redshift bins as indicated in the lower-left part of panel. Seven redshift bins are used, from $z=0$ to 3.5, centred on $z$=0.075, 0.28, 0.55, 0.85, 1.25, 1.85, and 2.85. It is notable that the faint end of RLFs begin to flatten visibly at redshift of $z\thicksim 1.25$. These results thus indicate a luminosity-dependent evolution for steep-spectrum radio sources.}
\end{figure*}

\subsection{The total RLF at 408 MHz}

By choosing seven redshift bins (indicated in the lower-left part of panel in Fig.4) and the luminosity bins of width $\Delta\log_{10}L_{408}=0.3$, the binned total RLF at 408 MHz is obtained and shown in Fig.4.

The binned RLF can be approximately described by broken power-law, although the case of redshift bin $0<z<0.15$ is a little special. Similar to luminosity functions at other wavebands \citep[see][]{b4,b34}, the RLF is steeper at the bright end and flatter at the faint end. This shape is rather remarkable at high redshift.

The low-luminosity (composed of FRIs and low-excitation/weak emission line FRIIs) and the high-luminosity sources are believed to have different evolution \citep{b11,b43}. The faintest sub-sample of our combined sample is the MRC 1-Jy sample with a flux limit of 0.95 Jy. We find that this sample is still too small to decide the evolution trends of low-luminosity sources. Especially at high redshift, the RLFs spread in a very narrow luminosity range, while at the lowest redshift bin $0<z<0.15$, the RLF shows a obvious two-population composition. The main reason is that the FR I sources are dominant at very low luminosity ($\thicksim \log_{10}(L_{408} / {\rm W Hz^{-1}})<25$), which is an approximate dichotomy luminosity of FR Is and FR IIs \citep{b14}). The RLF at redshift bin $0<z<0.15$ thus represents a combination of two populations. More detailed discussion on this problem will be the subject of our future work which deals with analytical description of the RLFs.

\begin{figure}
  \centerline{
    \includegraphics[scale=0.46,angle=0]{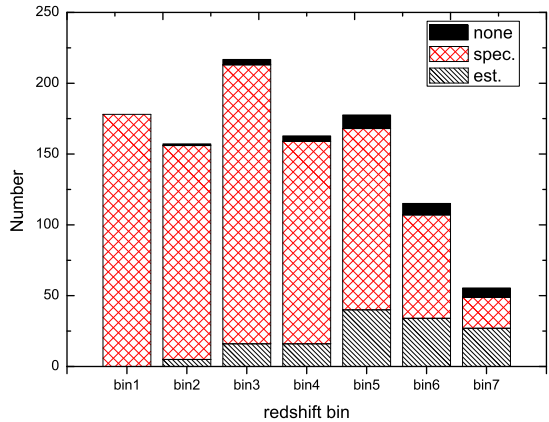}}
  \caption[redshift distribution]{\label{Photoz} The number distribution for the seven redshift bins. Sources with spectroscopic and $R-z$ or $K-z$ estimated redshifts are represented by the red cross-hatched and gray hatched columns. For the 33 sources without redshifts, we assume that they have the same redshift distribution as the 138 estimated ones. They are represented by black solid columns.}
\end{figure}

\subsubsection{Sources without spectroscopic redshifts}

In our combined sample, 171 sources have no spectroscopic redshifts. As shown in section 2.2, photometric redshifts of 138 sources are estimated, leaving still 33 sources without redshifts. The number distribution for the seven redshift bins (indicated in the lower-left part of panel in Fig.4) is shown in Fig.5. It can be noticed that the sources with estimated redshifts are mainly located in the three highest redshift bins. For the 33 sources without redshifts, which are mainly radio galaxies from the MRC 1-Jy sample, we assume that they have the same redshift distribution as the 138 estimated ones. Owing to intrinsic scatters (see Fig.1 and Fig.2) in the empirical relation of $R-z$ and $K-z$, more or less error will be introduced into the total RLFs. Especially for the last two redshift bins ($1.5<z<2.2$ and $2.2<z<3.5$), the percentage of sources with estimated redshifts are fairly high ($\thicksim$ 30\% and 50\%). Taking into account the intrinsic scatters in the two relations, the mean square error of redshift estimation is 0.4427. The error introduced into the RLFs for the last two redshift bins is grossly estimated as 0.4 order of magnitude, roughly comparable with the error given by Eq.(5).

\subsubsection{The redshift cut-off}
Based on our sample, we also study the `redshift cut-off' question, involving whether the comoving number density of radio AGNs declines dramatically at the redshift beyond $z\thicksim 2.5$. The RLFs of flat-spectrum radio sources support the existence of a redshift cut-off \citep[e.g.][]{b11,b37,b22,b41}, while the `redshift cut-off' in the steep-spectrum radio sources has much controversy. \citet{b11} firstly  proposed a `redshift cut-off' for steep-spectrum radio sources, while several authors \citep{b23,b9} subsequently refuted this conclusion. The RLF we obtained doses not show an obvious `redshift cut-off' over the whole luminosity range. However, as shown in above, the bias introduced by the 33 sources without redshifts and 138 ones with only estimated redshifts can not be excluded for the last two redshift bins. Considering these effects, the `redshift cut-off' is still not obvious in our RLFs.

\subsubsection{The luminosity-dependent evolution}

The faint end of the RLF begins to flatten at $z\thicksim 1.25$, implying that the positive density evolution of fainter sources slows down beyond $z\thicksim 1.25$. From the binning RLF at 2.7 GHz, \citet{b40} concluded that the comoving density of these sources begins to decline at $1\lesssim z\lesssim 2$, while the actual turnover redshift may depend on the radio luminosity, implying the evolution to be a function of the intrinsic power. Our results thus support the luminosity-dependent evolution and agree with other previous researches \citep{b39,b7}. In fact, the luminosity-dependent evolution is also confirmed by optical and X-ray selected AGNs \citep{b20,b10}.

\begin{figure}
  \centerline{
    \includegraphics[scale=0.46,angle=0]{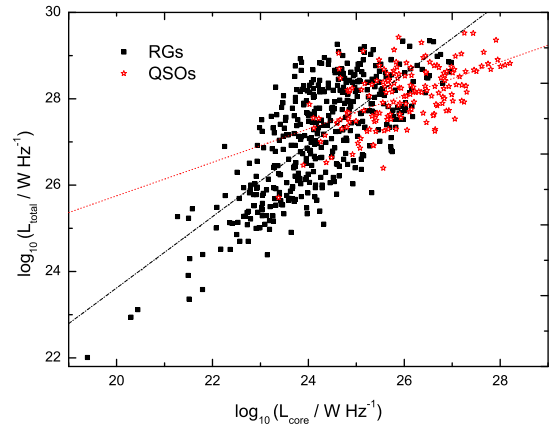}}
  \caption[R-z relation]{\label{Photoz}Relation of core luminosity at 5 GHz vs. total luminosity at 408 MHz. The black squares and red stars correspond to radio galaxies and QSOs populations respectively. The black dashed and red dot lines show the linear fits.}
\end{figure}

\subsection{The core-total power correlation}

The total versus core radio luminosity for our sample of radio sources is plotted in Fig.6. In agrement with previous results \citep[e.g.][]{b44,b27}, we also find a strong correlation for radio galaxies (RGs) and a weaker correlation for QSOs. The relations are  $\log_{10}L_{total}=\log_{10}L_{core}\times(0.83 \pm 0.04)+(7.10 \pm 0.90)$ for RGs and $\log_{10}L_{total}=\log_{10}L_{core}\times(0.39 \pm 0.04)+(18.00 \pm 1.15)$ for QSOs. The RGs have a steeper slope than QSOs, indicating that their total power increases more rapidly with core power. RGs span a wider region in the total power, while QSOs are mainly distributed in the bright region.

\subsection{The core RLF at 5.0 GHz}

As the most active parts of radio AGNs, the cores are compact and bright. There are no surveys to search radio cores exclusively. The core fluxes are often derived by high-resolution observation after their parent radio sources have been identified. Our samples of cores thus do not have a flux limit as their parent radio samples do. Seeing that the 3CRR sample is almost complete (99.4\%) for core flux densities, it is used separately to estimate the binned core RLF by the method introduced in section 3.1. The relevant minimum core flux density is used as flux limit.

\begin{figure}
  \centerline{
    \includegraphics[scale=0.46,angle=0]{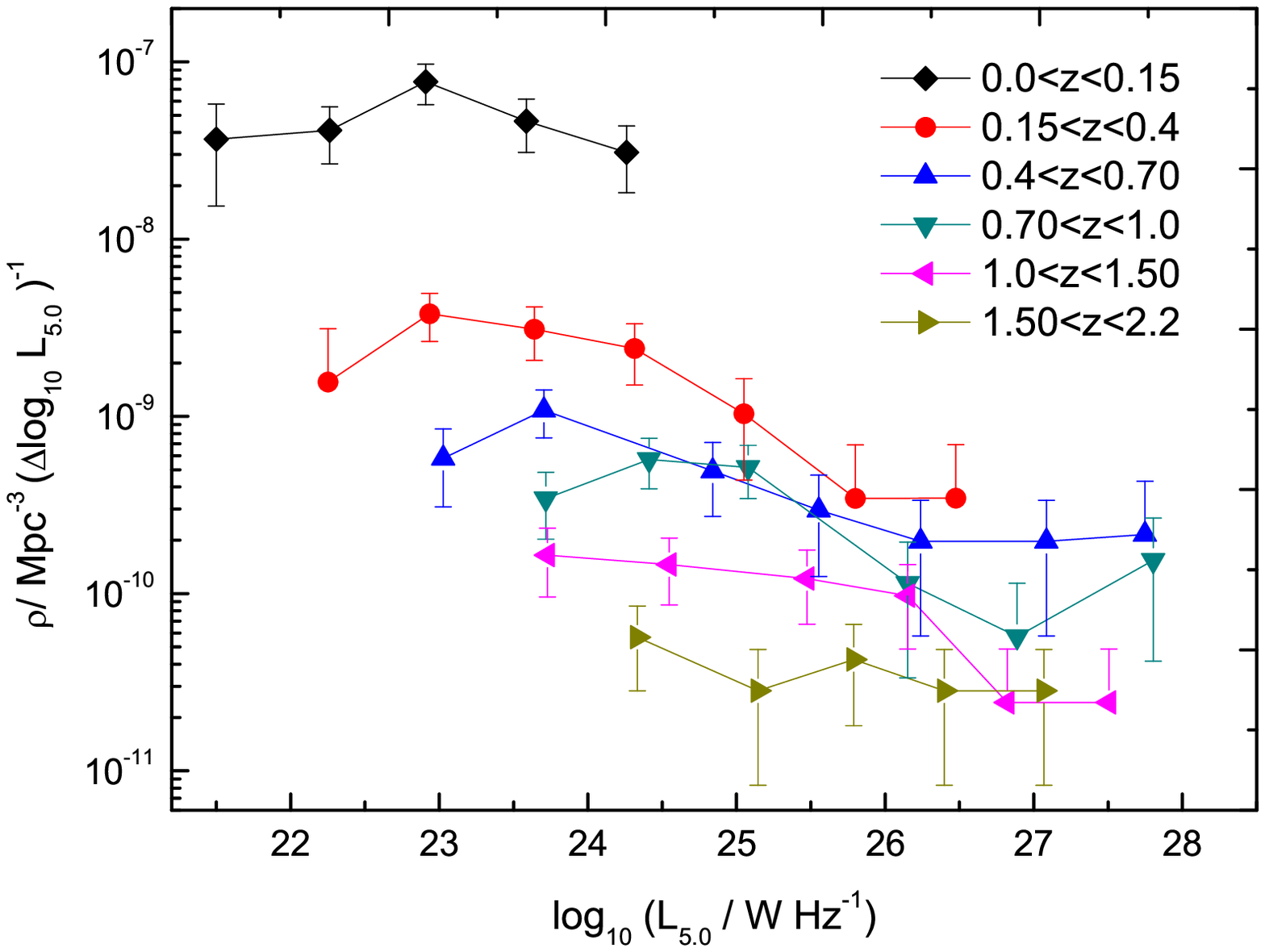}}
  \caption[R-z relation]{\label{Photoz}The binned core RLFs based on the 3CRR data are plotted with Poisson errors from Eq.(4). Different symbols correspond to different redshift bins as indicated in the upper-right part of panel.}
\end{figure}

\begin{figure}
  \centerline{
    \includegraphics[scale=0.46,angle=0]{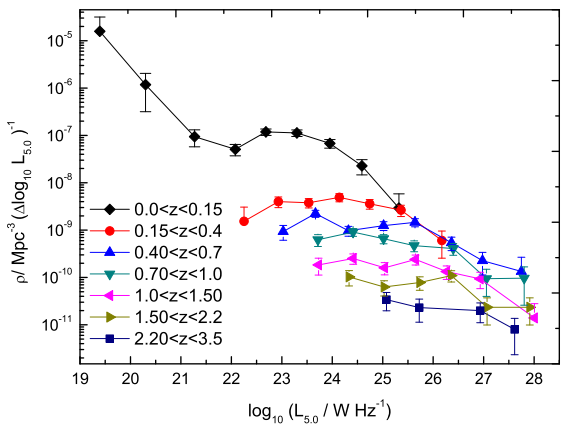}}
  \caption[R-z relation]{\label{Photoz}The binned core RLFs based on the combined sample are plotted with Poisson errors from Eq.(4). Different symbols correspond to different redshift bins as indicated in the lower-left part of panel.}
\end{figure}

\begin{figure*}
  \centerline{
    \includegraphics[scale=0.95,angle=0]{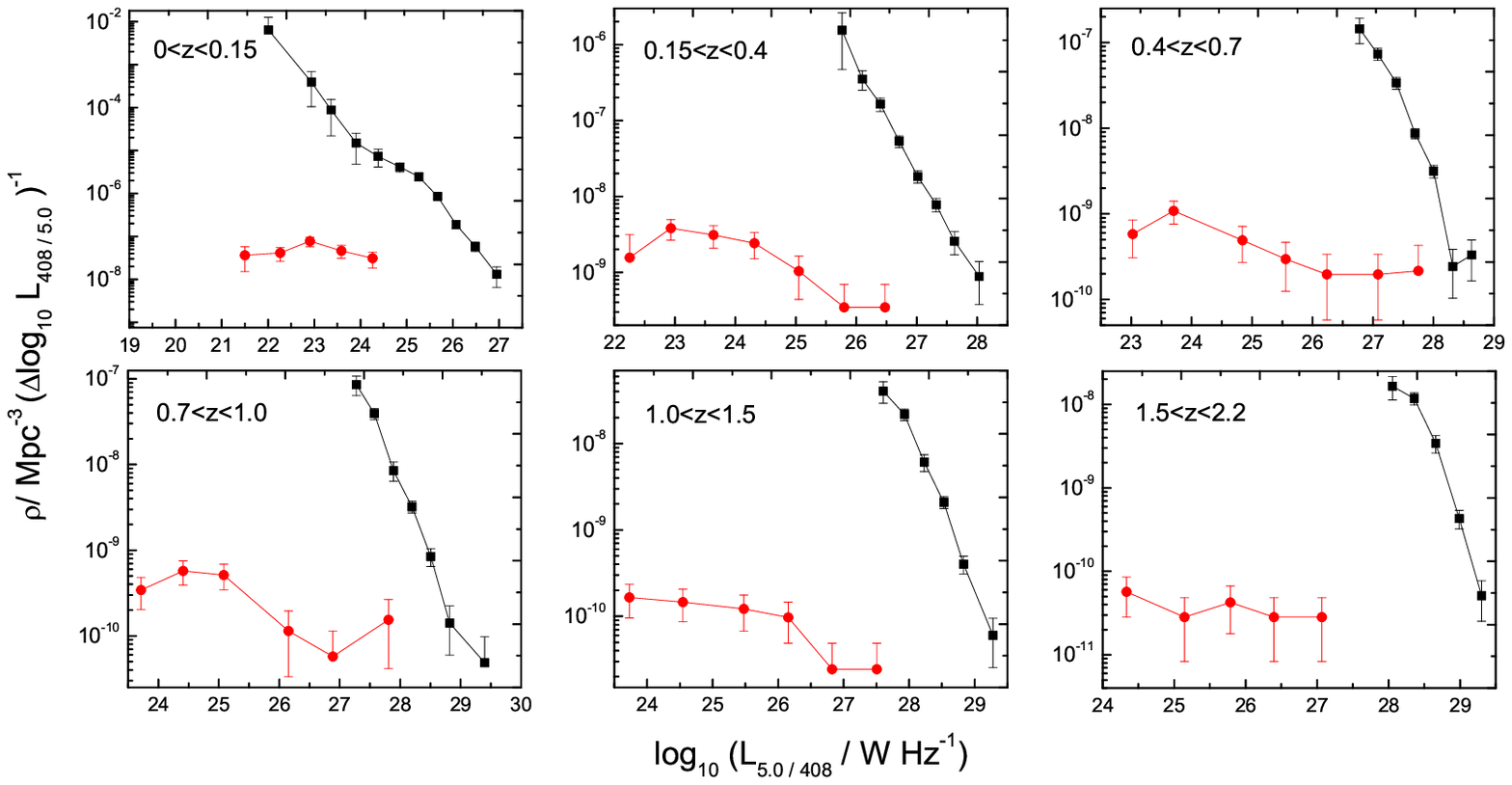}}
  \caption[R-z relation]{\label{Photoz}Comparison of core-total RLF for each redshift bin. Red solid circle and black rectangular correspond to the core RLF and total RLF respectively. Due to the different frequency of core RLF and total RLF (5 GHz and 408 MHz respectively), the term $L_{5.0/408}$ is used.}
\end{figure*}

Choosing six redshift bins (shown in the upper-right part of panel in Fig.7) and luminosity bins of width $\Delta\log_{10}L_{5.0}=0.6$, we obtain the binned core RLF at 5.0 GHz shown in Fig.7. The core RLF is obviously different from the total RLF. It is very flat at the whole luminosity range. Even if at the bright end, the core RLF does not steepen as the total RLF. The comoving number density of radio cores presents a persistent decline with redshift, implying a negative density evolution.

To make full use of the core flux density data, we also estimate the binned core RLF based on our combined sample. Owing to the incompleteness (56.5\%) of core fluxes in the sample, we define the modified factor of completeness $C_{j}$ for a sub-sample $j$, where $C_{j}$ is the number percentage of sources with both core fluxes and redshifts. By choosing seven redshift bins (indicated in the lower-left part of panel in Fig.8) and luminosity bins of width $\Delta\log_{10}L_{5.0}=0.6$, the binned core RLF is obtained and shown in Fig.8. Although this result is not robust due to the incompleteness of core fluxes, we present it here for reference. This result is consistent with the one given by the 3CRR sample, supporting the negative density evolution of radio cores.

\subsection{Comparing core RLF with total RLF}

In order to compare the core RLF from the 3CRR sample and the total RLF in more detail,  both RLFs are plotted together for all the redshift bins in Fig.9. The shapes of core and total RLFs are obviously different, implying the different evolution of their radio emission. Because the radio sources in our combined sample are all lobe-dominant ones, the above result indicate the radio lobes and cores might not be co-evolving.

\section[]{Discussion}

The main result of this paper is the strikingly different evolution of the `cores' of radio sources compared to their extended radio emission. It is already known that the number density of extragalactic radio sources with both steep-spectrum and flat-spectrum increase sharply from $z=0$ to $z=2$ \citep[e.g.][]{b41}, while our study appears to show negative evolution for the number density of radio cores. Especially for the flat-spectrum sources, they are believed to be dominated by cores. We argue that although the flat-spectrum sources are dominated by cores, they can not be regarded as pure cores after all. Furthermore, in the light of unified models of AGNs, the flat-spectrum sources have a jet oriented at a small angle with respect to the line of sight, and their radio core luminosity are boosted by jet beaming effect. This make the RLF of flat-spectrum sources more obscure. Thus we have no reason to believe that their evolution of number density must be in agreement with the radio cores.

\subsection{The radio cores}

The striking evolution of the radio cores induce us to reexamine what a radio core is. Thus a brief review of the radio cores is worthwhile. In the standard interpretation, the core is taken to be the optically thick base of the jet \citep{1979B}. This has been proved to be indeed the case as shown by VLBI maps \citep{2011A}. As \citet{FB1995} modeled, after the jet formed, the flow will be firstly accelerated in a nozzle close to the disk, then it will expand adiabatically, enclosing the magnetic field and showing up as the compact core. The radio core emission, which is generally thought to be self-absorbed nonthermal synchrotron emission originate in the inner jet \citep{2002V}, has a high degree of polarization, a high brightness temperature \citep{2008H}, and typically a flat spectral slope at radio wavelengths. Due to synchrotron self-absorption, the absolute position of the observed VLBI core shifts systematically with frequency, moving increasingly outward along the VLBI jet with lower frequency \citep{2011K}.

\subsection{The negative evolution of radio cores}

As shown in Fig.7 and Fig.8, the comoving number density of radio cores presents a persistent decline with redshift, implying a negative density evolution. It seems that the core radio emission could be gradually powered by central engines with cosmic epoch, or the radio-loudness of the cores be epoch-dependent. Specifically, the radio loudness of the cores may be weaker in earlier epoch, and fewer cores can be observed in radio emission. Indeed, amongst the galaxies harbouring an AGN, only a small fraction (about 10\%) appear to be luminous at radio wavelengths \citep{b10}, and there have been suggestions that the radio loudness may be epoch-dependent \citep[also see][]{SJ09}.

\citet{2002H} and \citet{2007S} found that there is a strong correlation between radio loudness and Eddington ratio ($L_{bol}/L_{Edd}$, the ratio of bolometric luminosity to Eddington luminosity), indicating that radio loudness decreases with increasing $L_{bol}/L_{Edd}$. Both studies used the $R$ parameter, defined as the ratio of the monochromatic 5 GHz radio luminosity to the 4400 {\AA} optical luminosity ($R\equiv L_{\nu5}/L_{\nu B}$), to characterize the degree of radio loudness. \citet{2007S} used the total luminosity (i.e., including nuclear and extended emission) to evaluate the radio loudness, while \citet{2002H} only took into account the nuclear emission. In spite of the difference, the trends of decreasing radio loudness with increasing Eddington ratio were consistent. \citep{1995F} argued that the difference between radio loud and radio weak is established already on the parsec scale. Therefore the above trends actually characterize the property of cores, implying the radio loudness of cores to be lower in early epochs. These trends are consistent with the property of core RLF we obtained.

The epoch-dependent radio loudness was also found by \citet{2007J}. Using a sample of more than 30,000 optically selected quasars from the Sloan Digital Sky Survey (SDSS), they found the radio-loud fraction of quasars is a strong function of redshift, i.e., the radio-loud fraction decreases rapidly with increasing redshift. They also used the $R$ parameter to characterize the degree of radio loudness, and chose $R > 10$ to define radio-loud quasars. It is shown that the radio loudness of quasars are weaker at higher redshifts, thus the fraction of radio loud quasars is lower in earlier epochs. Considering that most of the radio loud quasars in their sample are core-dominated FR I instead of FR II, their result that the radio-loud fraction of quasars negatively evolves with redshift is also in agreement with our result of core RLF.

Besides, as a good supplement to the above studies, \citet{2009R} investigated how the radio loud fraction of AGNs changed as a function of Eddington ratio. They found that 4.7\% of the AGNs in their flux-limited subsample are radio loud based on core radio emission alone. The radio loud fraction decreases from 13\% to 2\% as Eddington ratio increases over 2.5 orders of magnitude. This result is consistent with the above findings, and also supports the result of our core RLF.

\subsection{The analogy with XRBs}

Interestingly, it has become increasingly clear that many aspects of BH accretion and jet formation are directly comparable between AGNs
and X-ray binary (XRB) systems \citep[e.g.][]{2010F,2006K}. The XRBs may help one to understand the radio emission evolution of AGNs. Studies \citep[see][]{2004F,2010F} have shown that the jet power associated with radio emission, as well as the radiative efficiency of the accretion flow, can change dramatically among different accretion states for XRBs. In the so-called low/hard state, the Eddington ratio is very low (below about 0.01), and a compact self-absorbed jet with radio flat spectrum appears. With the arrival of higher Eddington ratios, sources can switch into `softer' states in which the radio emission is dramatically suppressed or even quenching. It implies a scenario of jet formation that the jet contributing radio emission is more easily produced in sources with lower accretion ratio. For AGNs, the Eddington ratios trend to be higher in early epochs and suppress the production of jet contributing radio emission, their core radio emission is then weaker at higher refshifts. When the Eddington ratios decrease with decreasing redshifts, AGNs easily produce the jet to contribute strong core radio emission. Our core RLF confirms that the jet is more easily formed in sources with lower accretion ratio.

\subsection{The different evolution of radio cores and lobes}

It is well known that the cores and the extended lobes are related in matter and energy by jet. The core-total power correlation (see Fig.6) is thus the immediate representation of this relation. Our results indicate that, however, they behave different evolution pattern and could not be co-evolving. We believe that some factors could affect their evolution connection. Firstly, the long distance between the core and lobe delays the exchange of matter and energy, and weaken their co-evolution. For a 100-kpc AGN jet, we see the core change up to 1 Myr earlier than the near lobe, assuming a jet velocity of 0.1c (c is the velocity of light). Besides, the lobes are inevitably affected by intergalactic environment, leading to the evolution uncertainty. The lobes expand in the intergalactic medium (IGM), while the cores are located in interstellar medium. Different environments (such as magnetic field and particle number density) will lead distinct energy loss processes \citep[e.g.][]{2009K}. Compared with lobes, the radio radiation properties of cores are more sensitive to the accretion states of central BHs. As enormous reservoirs of energy and relativistic electrons, the lobes, even if the radio cores have been quenching, may still remain detectable for a long time if they are subject only to radiative losses of the relativistic electrons \citep{M11}. Assuming a typical magnetic intensity of 1.5 nT \citep[1 nT = 10$\mu$G, see][]{2005C} of the lobes, the fading time of the lobes at 408 MHz is estimated as,
\begin{eqnarray}
t_{fa} \simeq t_{sy} = 6.0\times10^{7}(\frac{B}{nT})^{-3/2}(\frac{\nu}{0.5GHz})^{-1/2} yr
\end{eqnarray}
which is about $10^{7}$ years.

The episodic AGN activity could also affect the connection of evolution between lobes and cores. There has been a growing body of evidence to suggest that AGN activity could be episodic, although the range of time scales involved needs to be investigated further \citep[e.g.][]{B07,BKS11,SJ09}. During the phase of inactivity, the energy supply from core to extended regions ceases and a source is expected to undergo a period of fading before it disappears completely \citep{MS09}. In this case, sources may lack certain features, such as radio cores or well-defined jets that are produced by continuing activity. Comparatively, the radio lobes are not so sensitive to continuing activity than radio cores. Alternatively, during the phase of reactivation, one may expect to observe fossil radio lobes remaining from an earlier active epoch, along with newly restarting jets and cores \citep{M11}. In this case, the lobes are already disconnected from the currently active jets and cores.

\section[]{Conclusions}

We concentrate our efforts on the study of the correlation between radio galaxies/QSOs and their cores via radio luminosity functions. Using a large combined sample of 1063 radio-loud AGNs selected at low radio frequency, we investigate the radio luminosity function (RLF) at 408 MHz band. We also estimate the core RLF at 5 GHz band based on the 3CRR sample and the combined sample. Main results are follow as:

\begin{enumerate}
  \item In agrement with previous results, we note a strong correlation between core and total radio power for RGs and QSOs, but the correlations has large dispersion, especially for QSOs.  We find that the total power of RGs more strongly depend on core radio power compared to QSOs.
  \item Looking at the possible existence of a `redshift cut-off', the steep-spectrum RLFs we obtained do not show an obvious density decline for powerful radio sources beyond $z \thicksim 2.5$ over the whole luminosity range. We argue that the evolution of radio AGNs is luminosity-dependent, while the so-called `redshift cut-off' in steep-spectrum population is not obvious in our result of RLFs.
  \item The core RLFs we obtained show that the comoving number density of radio cores has a persistent decline with redshift, implying a negative density evolution. We believe that the radio core emission could be gradually powered by central engines, or their radio-loudness be epoch dependent.
  \item It is noticed that the core RLF is obviously different from the total RLF at 408 MHz band which is mainly contributed by extended lobes, implying that the core and extended lobes could not be co-evolving at radio emission.
\end{enumerate}

\section*{Acknowledgments}

We are grateful to the referee for very useful comments that improved this paper. We acknowledge the financial supports from the National Natural Science Foundation of China 10778702, the NationalBasic Research Program of China (973 Program 2009CB824800), and the Policy Research Program of Chinese Academy of Sciences (KJCX2-YW-T24). ZLY thanks Professor P. J. McCarthy for offering K-band magnitude data of the MRC 1-Jy sample, Dr. C. De Breuck and Wil van Breugel for some useful help. We thank Hardcastle team providing the 3CRR data on the Web site. This research has made use of the NASA/IPAC Extragalactic Database (NED) which is operated by the Jet Propulsion Laboratory, California Institute of Technology, under contract with the National Aeronautics and Space Administration.

\appendix

\section{The Combined Sample}\label{details}

The list of combined sample of radio galaxies/QSOs. The table has been arranged as follows:
\medskip

  Column (1). Source name in IAU designation (B1950). An asterisk following the source name indicates that the source should be excluded from the combined sample for the reason explained in section 2. They are put in the table just for reference.

  Column (2). Other name if available.

  Column (3). Redshift. The redshift with a suffix e is estimated by the empirical relation of $R-z$ or $K-z$ as mentioned in section 2.2. Otherwise, redshifts are spectroscopic.

  Column (4). Total flux density at 408 MHz in Jy. For the 3CRR sample, these fluxes are from literature or extrapolated from 178 MHz fluxes.

  Column (5). Spectral index $\alpha$ near 408 MHz (where $S=\nu^{-\alpha}$).

  Column (6). Core flux density at 5 GHz in mJy.

  Column (7). Core spectral index $\alpha_{c}$ near 5 GHz. Generally, core spectral index information is far from abundant in literature. Most of our sources with core fluxes do not have core spectral indices. For these sources, we simply assume a flat spectrum, $\alpha_{c} =0$, in order to calculate core luminosities.

  Column (8). References. For the 3CRR sample, these are mainly references for total and core fulx densities as well as core spectral indices; for other samples, these are mainly references for redshifts, core flux densities and core spectral indices. The references for total flux densities at 5 GHz, which are used to calculate the $R_{c}$-parameter (see section 2.5), are also arranged in this column.

  Column (9). The sub-sample name. e.g. 3CRR: the 3CRR sample; MS4: the Molonglo southern 4 Jy sample; MRC1: the MRC 1-Jy sample; BRL: the BRL sample.

\begin{deluxetable}{cllrcrrcc}
\tablecolumns{9}
\tabletypesize{\footnotesize}
\tablewidth{0pt}
\tablecaption{The Combined Sample}
\tablehead{
\colhead{IAU}   &
\colhead{Other}  &
\colhead{$z$}  &
\colhead{$S_{t0.408}$}  &
\colhead{$\alpha$} &
\colhead{$S_{core5.0}$} &
\colhead{$\alpha_{c}$}      &
\colhead{Ref.}      &
\colhead{Sample} \\
\colhead{}   &
\colhead{Name}   &
\colhead{}   &
\colhead{Jy}   &
\colhead{}      &
\colhead{mJy}  &
\colhead{} &
\colhead{} &
\colhead{}
}

\startdata
0007$+$124  &  4C12.03  &  0.156       &   4.45     &  0.87        &       3.5       &             &  1,54,55,57         &  3CRR    \\
0013$+$790  &  3C6.1    &  0.8404      &   8.70     &  0.68        &       4.4       &             &  4,54,55,57         &  3CRR    \\
0038$-$294  &           & 3.0708$_{e}$ &   1.01     &  0.87        &       41.1      &             &  15,41,11           &  MRC1    \\
0017$-$207  &           & 0.545        &   1.25     &  0.95        &       2.8       &     0.09    &  18,42,28           &  MRC1    \\
0003$-$567  &           &  0.2912      &  5.19      &  0.83        &                 &             &  7,40               &  MS4     \\
0003$-$428  &           &  0.53        &   5.15     &  0.86        & $<$   143       &             &  7,40               &  MS4     \\
0003$-$003  &  3C2      &  1.037       &  10.45     &  0.80        &       48        &     0       &  19,6               &  BRL     \\
0034$-$014  &  3C15     &  0.073       &   9.74     &  0.71        &       299       &     -0.28   &  12,13,19           &  BRL     \\
\enddata
\tablenotetext{a}{Table 1 is available in its entirety in machine-readable forms in the online journal.
A portion is shown here for guidance regarding its form and content.}
\tablerefs{ (1) \citet{PKS90}; (2) \citet{a19}; (3) \citet{a35}; (4) \citet{b17}; (5) \citet{a18}; (6) \citet{a49}; (7) \citet{b6}; (8) \citet{a16}; (9) \citet{a52}; (10) \citet{a30}; (11) \citet{a50}; (12) \citet{a44}; (13) \citet{a45}; (14) \citet{a8}; (15) \citet{b25}; (16) \citet{a41}; (17) \citet{a26}; (18) \citet{b24}; (19) \citet{b2}; (20) \citet{a6}; (21) \citet{a5}; (22) \citet{a38}; (23) \citet{a17}; (24) \citet{a3}; (25) \citet{a1}; (26) \citet{a20}; (27) \citet{a40}; (28) \citet{a28}; (29) \citet{a27}; (30) \citet{a34}; (31) \citet{a46}; (32) \citet{a54}; (33) \citet{a53}; (34) \citet{a56}; (35) \citet{a13}; (36) \citet{a14}; (37) \citet{a22}; (38) \citet{a25}; (39) \citet{a48}; (40) \citet{b5}; (41) \citet{b33}; (42) \citet{a2}; (43) \citet{a11}; (44) \citet{a24}; (45) \citet{a43}; (46) \citet{a29}; (47) \citet{a15}; (48) \citet{a7}; (49) \citet{a12}; (50) \citet{a55}; (51) \citet{a51}; (52) \citet{a39}; (53) \citet{a31}; (54) \citet{b26}; (55) \citet{b19}; (56) \citet{a37}; (57) \citet{b35}; (58) \citet{B98}; (59) \citet{H97}; (60) \citet{G04}; (61) \citet{b18}; (62) \citet{D08}; (63) \citet{M2005}; (64) \citet{2006T}; (65)\citet{1980L}; (66)\citet{1991B}}
\end{deluxetable}

\end{document}